\documentclass[12,article]{emulateapj}

\shorttitle{Optically Dull AGNs}
\shortauthors{Trump et~al.}

\begin{document}

\def\etal{et al.}

\title{The Nature of Optically Dull Active Galactic Nuclei in COSMOS\altaffilmark{1}}

\author{
  Jonathan R. Trump,\altaffilmark{2}
  Chris D. Impey,\altaffilmark{2}
  Yoshi Taniguchi,\altaffilmark{3}
  Marcella Brusa,\altaffilmark{4}
  Francesca Civano,\altaffilmark{5}
  Martin Elvis,\altaffilmark{5}
  Jared M. Gabor,\altaffilmark{2}
  Knud Jahnke,\altaffilmark{6}
  Brandon C. Kelly,\altaffilmark{5}$^,$\altaffilmark{7}
  Anton M. Koekemoer,\altaffilmark{8}
  Tohru Nagao,\altaffilmark{3}
  Mara Salvato,\altaffilmark{9}
  Yasuhiro Shioya,\altaffilmark{3}
  Peter Capak,\altaffilmark{9}
  John P. Huchra,\altaffilmark{5}
  Jeyhan S. Kartaltepe,\altaffilmark{10}
  Giorgio Lanzuisi,\altaffilmark{11}
  Patrick J. McCarthy,\altaffilmark{12}
  Vincenzo Maineri,\altaffilmark{13}
  and Nick Z. Scoville\altaffilmark{9}
}

\altaffiltext{1}{ 
  Based on observations with the NASA/ESA \emph{Hubble Space
  Telescope}, obtained at the Space Telescope Science Institute, which
  is operated by AURA Inc, under NASA contract NAS 5-26555; the
  Magellan Telescope, which is operated by the Carnegie Observatories;
  and the Subaru Telescope, which is operated by the National
  Astronomical Observatory of Japan.
\label{cosmos}}

\altaffiltext{2}{
  Steward Observatory, University of Arizona, 933 North Cherry
  Avenue, Tucson, AZ 85721
\label{Arizona}}

\altaffiltext{3}{
  Research Center for Space \& Cosmic Evolution, Ehime University, 2-5
  Bunkyo-cho, Matsuyama 790-8577, Japan
\label{Ehime}}

\altaffiltext{4}{
  Max Planck-Institut f\"ur Extraterrestrische Physik,
  Giessenbachstrasse 1, D-85748 Garching, Germany
\label{Max Planck}}

\altaffiltext{5}{
  Harvard-Smithsonian Center for Astrophysics, 60 Garden Street,
  Cambridge, MA 02138
\label{CfA}}

\altaffiltext{6}{
  Max Planck Institut f\"ur Astronomie, K\"onigstuhl 17, D-69117
  Heidelberg, Germany
\label{Max Planck 2}}

\altaffiltext{7}{
  Hubble Fellow
\label{Hubble}}

\altaffiltext{8}{
  Space Telescope Science Institute, 3700 San Martin Drive, Baltimore,
  MD 21218
\label{Maryland}}

\altaffiltext{9}{
  California Institute of Technology, MC 105-24, 1200 East California
  Boulevard, Pasadena, CA 91125
\label{Caltech}}

\altaffiltext{10}{
  Institute for Astronomy, 2680 Woodlawn Dr., University of Hawaii,
  Honolulu, HI 96822
\label{Hawaii}}

\altaffiltext{11}{
  Dipartimento di Fisica, Universit\`a di Roma La Sapienza, P.le
  A. Moro 2, 00185 Roma, Italy
\label{Rome}}

\altaffiltext{12}{
  Observatories of the Carnegie Institute of Washington, Santa Barbara
  Street, Pasadena, CA 91101
\label{Carnegie}}

\altaffiltext{13}{
  European Southern Observatory, Karl-Schwarschild-Strasse 2, D-85748
  Garching, Germany
\label{ESO}}

\newcommand{\OII}{\hbox{{\rm [O}\kern 0.1em{\sc ii}]}}
\newcommand{\OIII}{\hbox{{\rm [O}\kern 0.1em{\sc iii}]}}

\begin{abstract}

We present infrared, optical, and X-ray data of 48 X-ray bright,
optically dull AGNs in the COSMOS field.  These objects exhibit the
X-ray luminosity of an active galactic nucleus (AGN) but lack broad
and narrow emission lines in their optical spectrum.  We show that
despite the lack of optical emission lines, most of these optically
dull AGNs are not well-described by a typical passive red galaxy
spectrum: instead they exhibit weak but significant blue emission like
an unobscured AGN.  Photometric observations over several years
additionally show significant variability in the blue emission of four
optically dull AGNs.  The nature of the blue and infrared emission
suggest that the optically inactive appearance of these AGNs cannot be
caused by obscuration intrinsic to the AGNs.  Instead, up to
$\sim$70\% of optically dull AGNs are diluted by their hosts, with
bright or simply edge-on hosts lying preferentially within the
spectroscopic aperture.  The remaining $\sim$30\% of optically dull
AGNs have anomalously high $f_X/f_O$ ratios and are intrinsically
weak, not obscured, in the optical.  These optically dull AGNs are
best described as a weakly accreting AGN with a truncated accretion
disk from a radiatively inefficient accretion flow.

\end{abstract}


\section{Introduction}

Deep X-ray surveys have indicated that most X-ray sources in the sky
are AGNs with a wide range of luminosities, spectral energy
distributions (SEDs), and redshifts \citep[e.g. ][]{bru07, luo08,
ued08}.  X-ray selection is widely regarded as the most efficient
method for finding AGNs \citep{ris04, bra05} and most of the X-ray
background has been resolved into discrete AGN point sources
\citep[e.g. ][]{ale03, bau04, bal07}.  Most X-ray selected AGNs are
quite similar to bright quasars from optical surveys, but many would
not be easily selected as AGNs by their optical emission.  The class
of ``optically dull'' AGNs \citep[also called ``X-ray bright,
optically normal galaxies,'' or XBONGs,][]{com02} are particularly
puzzling because their X-ray emission is bright even while the optical
signature of an AGN is completely absent.  First pointed out by
\citet{elv81}, optically dull AGNs lack both the broad emission lines
of unobscured Type 1 AGNs and the narrow emission lines of moderately
obscured Type 2 AGNs.  They are also different from heavily obscured
($N_H \gtrsim 10^{24}$ cm$^{-2}$) ``Compton-thick'' AGNs, which lack
both optical and X-ray emission and are frequently missed by X-ray
surveys.

What causes an optically dull AGN to have the bright X-ray emission of
an AGN while lacking all optical signatures of AGN accretion?  The
simplest possibility is that optically dull AGNs aren't special at
all, but are normal AGNs diluted by bright hosts.  \citet{mor02} in
particular suggest that local Seyfert galaxies would be classified as
optically dull if they were observed with large apertures (as is the
case at higher redshift, where the host galaxy is an unresolved source
fully within the spectroscopic slit or fiber).  However, 10-20\% of
local (undiluted) AGNs are optically dull \citep{laf02, hor05}, so
dilution may not be the cause of all optically dull AGNs.

Another possibility is that the optical emission of optically dull
AGNs is absorbed.  Narrow emission line (Type 2) AGNs have been long
thought to be Type 1 AGNs with an obscured broad line region
\citep[e.g.,][]{ant93}, and optically dull AGNs may similarly have the
entire narrow line region obscured.  \citet{com02} and \citet{civ07}
suggest gas and dust with a large covering fraction a few parsecs from
the nuclear source could provide the necessary absorption, blocking
the ionizing radiation from exciting the narrow line region.
\citet{rig06} instead suggest that optically dull AGNs are obscured by
extranuclear ($>$100 pc) gas and dust in the host galaxy.  No matter
the source of the gas and dust, obscuring optically dull AGNs would
require material which preferentially absorbs the optical emission,
since at least half of optically dull AGNs are relatively unobscured
($N_H<10^{22}$ cm$^{-2}$) in the X-rays \citep{sev03, pag03}.

Optically dull AGNs may instead be exotic AGNs with unusual emission
or accretion properties.  In particular, \citet{yua04} suggest that
optically dull AGNs may be radiatively inefficient accretors with
truncated accretion disks.  In this scenario, gas near the AGN does
not form a cool disk, but instead is a very hot, radiatively
inefficient, accretion flow (RIAF, also called an advection dominated
accretion flow, or ADAF).  This gas would then glow brightly in X-rays
from inverse Compton emission while lacking the optical/UV blackbody
emission from a typical AGN accretion disk.  RIAFs have been shown to
explain local low-luminosity AGNs \citep{qua99, shi00, nag02, hop09}.

We use a sample of 48 optically dull AGNs from the Cosmic Evolution
Survey \citep[COSMOS,][]{sco07}\footnote{The COSMOS website is
http://cosmos.astro.caltech.edu/.} to test these hypotheses.  We
describe the selection and multiwavelength observations in \S 2.  In
\S 3 we use a combination of photometry and spectroscopy to fit the
optical emission of the optically dull AGNs, revealing that most of
our targets show distinct contributions in the optical emission from a
weak blue AGN and a dominant red passive galaxy.  \S 3 also shows that
at least four of the optically dull AGNs show significant variability.
In \S 4 we summarize our findings and present the case that $\sim$70\%
of optically dull AGNs are normal AGNs diluted by their host galaxies,
while the remaining $\sim$30\% are instrinsically weak with
radiatively inefficient accretion.  We examine the accretion
properties of weak AGNs in detail in \S 5, and summarize our results
in \S 6.

\section{Observations}

\subsection{X-ray Selection}

We draw the sample of optically dull AGNs from the Cosmic Evolution
Survey \citep[COSMOS,][]{sco07}, a survey over 2 deg$^2$ of sky with
deep multiwavelength observations.  The XMM-{\it Newton} observations
of COSMOS reach limiting fluxes of 1.7$\times 10^{-15}$
erg~cm$^{-2}$~s$^{-1}$ and 9.3$\times 10^{-15}$ erg~cm$^{-2}$~s$^{-1}$
in the 0.5-2 keV and 2-10 keV energy bands, respectively
\citep{capp09}.  The optical and infrared counterparts to the X-ray
sources is presented in \citet{bru09}, and all counterpart matches
were visually inspected.  Spectroscopic follow-up of X-ray targets
with $i_{\rm AB} \le 23$ is described in \citet{tru09a}.  In
particular, the 48 optically dull AGNs are the objects of
\citet{tru09a} classified as ``a'' types (absorption line spectra)
with 90\% redshift confidence.  All objects lack strong emission lines
(see \S 2.2 below) in the optical spectra and satisfy one of the two
X-ray AGN criteria:

\begin{equation}
  L_{0.5-10~{\rm keV}} > 3 \times 10^{42}~{\rm erg~s}^{-1}
\end{equation}
\begin{equation}
  -1 \le X/O \le 1
\end{equation}

In Equation 2, $X/O = \log{f_X/f_O} = \log(f_{0.5-2~{\rm keV}}) +
i_{\rm AB}/2.5 + 5.352$.  These constraints are set by the limit on
X-ray luminosity in local star forming galaxies of $L_X \lesssim
10^{42}~{\rm erg~s}^{-1}$ \citep[e.g.,][]{fab89, col04} and the
traditional ``X-ray AGN locus'' of \citet{mac88}.  These equations
have been shown to be quite reliable in selecting AGNs, although they
are probably overly conservative \citep[e.g.,][]{hor01, ale01, bau04,
bun07}.  Of the 48 optically dull AGNs, 44 meet both criteria, with
only 4 meeting one criterion but not the other.  The optically dull
AGNs are additionally restricted to $z \le 1$, since beyond these
redshifts the 4000\AA~break shifts beyond the observed spectral range
of \citep{tru09a} and it becomes extremely difficult to measure
redshifts from absorption lines.  The 48 optically dull AGNs are all
of the $z<1$ AGNs within the 2 deg$^2$ of COSMOS that meet either of
the X-ray criteria and have Magellan/IMACS or SDSS spectroscopy.

\subsection{Spectroscopy}

Of the optically dull AGNs, 45/48 have optical spectra from
observations with the Inamori Magellan Areal Camera and Spectrograph
\citep[IMACS,][]{big98} on the 6.5 m Magellan/Baade telescope.  These
spectra have wavelength ranges of 5600-9200\AA, with a resolution
element of 10\AA~(5 pixels).  All targets were selected as AGN
candidates by their X-ray emission.  Details of the observations and
reductions are presented by \citet{tru09a}, and all of the spectra are
publicly available on the COSMOS IRSA server
(http://irsa.ipac.caltech.edu/data/COSMOS).  The optically dull AGNs
in this work all have high-confidence redshifts ($z_{\rm conf} \ge
3$), which empirically corresponds to a 90\% likelihood of the correct
redshift measurement \citep{tru09a}.

Three of the optically dull AGN spectra come from archival Sloan
Digital Sky Survey \citep[SDSS,][]{york00} observations.  These
sources were selected by their X-ray emission, but were excluded from
the main Magellan/IMACS survey because their redshifts were already
known.  Their wavelength coverage is {3800-9200\AA} and their
resolution element is {3\AA} (3 pixels).

\begin{figure}[ht]
\scalebox{1.2}
{\plotone{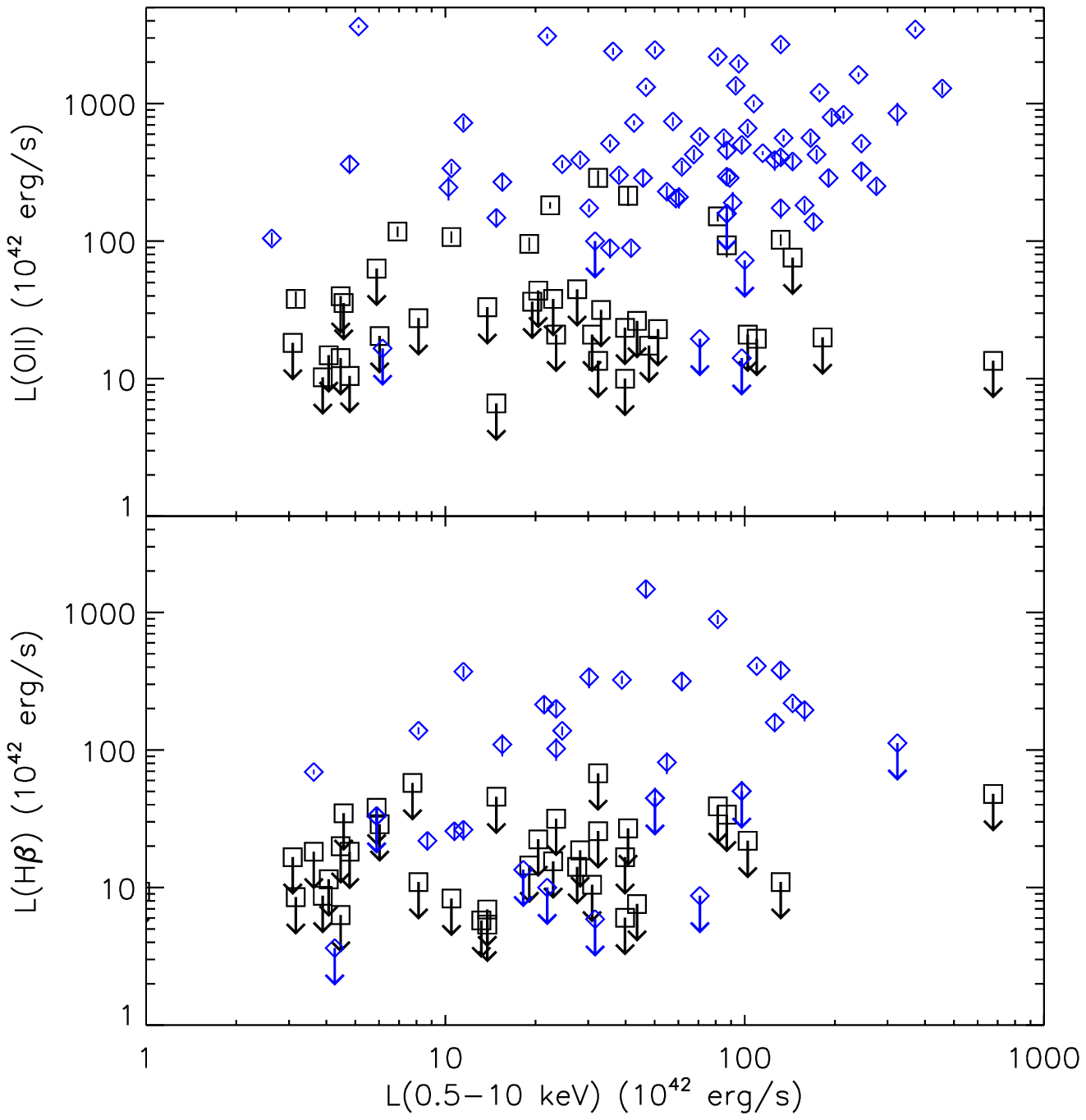}}
\caption{Luminosities of the {\OII}3727{\AA} and H$\beta$4861{\AA}
lines with X-ray luminosity for optically dull AGNs (squares) and Type
2 AGNs \citep[blue diamonds, from]{tru09a}.  For almost all optically
dull AGNs, no line is detected above the 5$\sigma$ threshold, and we
assign a 5$\sigma$ upper limit to the line luminosity.  Even when
$L>5\sigma_L$ and the line is positively measured, the {\OII} and
H$\beta$ lines are 10-100 times weaker than those of typical Type 2
AGNs.  Optically dull AGNs are not low-S/N Type 2 AGNs, but instead
have much less luminous emission lines despite their X-ray brightness.
\label{fig:linelimits}}
\end{figure}

Figure \ref{fig:linelimits} shows the measured {\OII}
($\lambda$3727\AA) and H$\beta$ ($\lambda$4861\AA) narrow emission
line luminosities for the optically dull AGNs (black squares), along
with a comparison sample of Type 2 AGNs (blue diamonds) from
\citet{tru09a}.  To compute the line luminosity, we first define a
straight-line continuum by averaging the spectral regions {30-40\AA}
redward and blueward of the line region.  The line luminosity is then
measured across the continuum-subtracted region 1000 km/s about the
line center.  The 1000 km/s width is a conservative limit since $<1\%$
of Type 2 AGNs have narrow emission lines broader than 1000 km/s
\citep{hao05}.  The measured error for each line luminosity is
computed using both the spectral error and the error of the continuum
fit.  When the line luminosity was less than its 5$\sigma$ error, we
used the 5$\sigma$ error as an upper limit on line luminosity.

All 38 optically dull AGNs with H$\beta$ in the observed wavelength
range have $L_{H\beta}<5\sigma_{L(H\beta)}$ and thus are assigned only
upper limits in the bottom panel of figure \ref{fig:linelimits}.
However, 9/40 optically dull AGNs with {\OII} in the observed
wavelength range have a line luminosity greater than the 5$\sigma$
threshold, despite the fact that the classification of \citet{tru09a}
identified no emission lines.  Still, even when detected, the emission
line luminosities of the optically dull AGNs are much lower than those
of typical Type 2 AGNs.  If the optically dull AGNs were simply Type 2
AGNs observed at low signal-to-noise (S/N), we might expect poorly
constrained upper limits on line luminosity.  This is not the case, as
the upper limits are 10-100 times lower than the line luminosities of
typical Type 2 AGNs.  Optically dull AGNs are not Type 2 AGNs with low
S/N, but have much less luminous emission lines than Type 2 AGNs of
similar X-ray luminosities.

\subsection{Optical and Infrared Photometry}

The optical and infrared photometry of the optically dull AGN is drawn
from the catalog of \citet{cap09}.  Table \ref{tbl:photometry} shows
the depths, wavebands, and year of observation for the COSMOS
photometry used here.

The optical photometry was taken on the Subaru telescope, with
observations of the 6 broad, 12 intermediate, and 2 narrow bands
described in \citet{tan07} and \citet{tan09}.  Fluxes were measured in
$3{\farcs}0$ diameter apertures, on PSF-matched images with FWHM of
$1{\farcs}5$, and simulations \citep{cap07} show that the $3{\farcs}0$
diameter aperture contains 76\% of the total flux for a point source.
We additionally correct each optical magnitude by the zero-point
correction from \citet{ilb09}.

The infrared photometry is derived from {\it Spitzer}/IRAC
observations.  The closest IRAC source within $1{\farcs}0$ of the
optical counterpart to the XMM source was chosen as the infrared
counterpart.  IRAC fluxes are given in the COSMOS-IRAC catalog for
$3{\farcs}8$ diameter apertures, so we translate these into
$3{\farcs}0$ diameter aperture fluxes as described in \citet{sal09}.
All of our optically dull AGNs were unambiguously detected in all 4
IRAC bands.

\begin{deluxetable}{cccccc}
\tablecolumns{6}
\tablewidth{0pc}
\tablecaption{COSMOS Optical and Infrared Photometry
\label{tbl:photometry}}
\tablehead{
  \colhead{Filter} &
  \colhead{Telescope} & 
  \colhead{Center $\lambda$} & 
  \colhead{FWHM} &
  \colhead{Depth ($3\farcs$0)} &
  \colhead{Epoch} \\
  \colhead{} &
  \colhead{} &
  \colhead{\AA} &
  \colhead{\AA} &
  \colhead{mag$_{AB}$} &
  \colhead{UTC} }
\startdata
B$_J$ & Subaru & 4460 & 897  & 27.7 & 2004 \\
g$^+$ & Subaru & 4750 & 1265 & 27.1 & 2005 \\
V$_J$ & Subaru & 5484 & 946  & 27.0 & 2004 \\
r$^+$ & Subaru & 6295 & 1382 & 27.1 & 2004 \\
i$^+$ & Subaru & 7640 & 1497 & 26.7 & 2004 \\
z$^+$ & Subaru & 9037 & 856  & 25.7 & 2004 \\
IA427 & Subaru & 4271 & 210  & 26.5 & 2006 \\
IA464 & Subaru & 4636 & 227  & 26.0 & 2006 \\
IA484 & Subaru & 4842 & 227  & 26.5 & 2007 \\
IA505 & Subaru & 5063 & 232  & 26.2 & 2006 \\
IA527 & Subaru & 5272 & 242  & 26.5 & 2007 \\
IA574 & Subaru & 5743 & 271  & 26.2 & 2007 \\
IA624 & Subaru & 6226 & 299  & 26.3 & 2006 \\
IA679 & Subaru & 6788 & 336  & 26.1 & 2006 \\
IA709 & Subaru & 7082 & 318  & 26.3 & 2007 \\
IA738 & Subaru & 7373 & 322  & 26.1 & 2007 \\
IA767 & Subaru & 7690 & 364  & 25.9 & 2007 \\
IA827 & Subaru & 8275 & 364  & 25.8 & 2006 \\
NB711 & Subaru & 7126 & 73   & 25.4 & 2006 \\
NB816 & Subaru & 8150 & 119  & 26.1 & 2005 \\
IRAC1 & Spitzer & 35263 & 7412  & 23.9 & 2006 \\
IRAC2 & Spitzer & 44607 & 10113 & 23.3 & 2006 \\
IRAC3 & Spitzer & 56764 & 13499 & 21.3 & 2006 \\
IRAC4 & Spitzer & 77030 & 28397 & 21.0 & 2006 \\
\enddata
\end{deluxetable}

\subsection{Host Morphologies}

We discuss morphological data of the optically dull AGN host galaxies
from observations with the Advanced Camera for Surveys (ACS) on the
{\it Hubble Space Telescope} (HST), fully described in \citet{koe07}.
The COSMOS field was imaged in the F814W filter for 583 orbits,
reaching a limiting magnitude of AB(F814W)$\le 27.2$ ($5\sigma$).
Because the HST/ACS imaging only covers 1.64 deg$^2$ of the 2 deg$^2$
COSMOS field, 3/48 optically dull AGNs lack HST/ACS coverage.
\citet{gab09} provides morphological data for 37 of the remaining
optically dull AGNs, from the point source subtracted host galaxies.
(The other 8 optically dull AGNs have HST/ACS imaging, but do not have
morphological data because the resultant fit was wildly unphysical or
did not converge.)



\subsection{Completeness}

All of the optically dull AGNs are detected in the Subaru and IRAC
photometry, and all within the HST areal coverage were detected in
ACS, so these do not affect the completeness limits.  The soft 0.5-2
keV X-ray limit of 1$\times 10^{-15}$ erg~cm$^{-2}$~s$^{-1}$ means the
X-ray AGN sample is complete to all AGNs meeting the luminosity
criterion (Equation 1) of $L_{0.5-10~{\rm keV}} > 3 \times
10^{42}~{\rm erg~s}^{-1}$ at $z \lesssim 1$ \citep[see Figure 9
of][]{tru09a}.  Correct identification of optically dull AGNs is also
limited to $z \lesssim 1$, since at higher redshifts the $4000\AA$
break in these objects is redshifted beyond the observed wavlength
range, and high-confidence identification becomes difficult.  The
Magellan/IMACS spectroscopy is uniformly 90\% complete to $i_{\rm AB}
\le 22$ absorption line objects.  The optically dull AGN sample is
then limited by $z \lesssim 1$ and $i_{\rm AB} \le 23$, but is 90\%
complete to only $i_{\rm AB} \le 22$.

\section{Multiwavelength Properties}

Table \ref{tbl:odagn} presents the multiwavelength properties of all
48 optically dull AGNs.  For each object, we show:
\begin{enumerate}
  \item The object name, with coordinates given in J2000
    hhmmss.ss+ddmmss.s.  ``COSMOS'' or ``SDSS'' indicates if the
    spectroscopy is from Magellan/IMACS or the SDSS archives,
    respectively.
  \item The redshift, from \citet{tru09a}.
  \item The signal to noise per pixel, averaged over the spectrum in
    the central wavelength range 6600-8200\AA.  (The resolution
    element is 5 pixels for Magellan/IMACS spectra, and 3 pixels for
    SDSS spectra.)
  \item The $i$-band AB magnitude, from Subaru/Suprime-Cam
    observations.
  \item The logarithm of the X-ray luminosity measured in the 0.5-10
    keV energy range, in cgs units.
  \item The ratio between X-ray and optical flux, $X/O$, defined in
    equation 2.
  \item The fractional contribution of AGN in the best-fit template
    (see \S 3.1).  This can be regarded as a rough estimate of the
    blue AGN contribution to the optical emission.
  \item The ratio between X-ray and optical flux, $X/O$, but where the
    optical flux includes only the blue AGN contribution (from the
    template fit in \S 3.1).
  \item The hardness ratio, $HR=(H-S)/(H+S)$.  Here $S$ is the flux in
    the soft 0.5-2 keV band and $H$ is the flux in the hard 2-10 keV
    band.  AGNs undetected in the soft band have $HR=1$, and those
    undetected in the hard band have $HR=-1$.
  \item The axis ratio $b/a$, measured using GALFIT after subtracting
    a point source from the host galaxy (see \S 3.5).
\end{enumerate}

\subsection{Optical Fitting: Host and AGN Components}

Each optically dull AGN spectrum lacks strong emission lines and has
the red shape and absorption signature (H+K lines, 4000\AA~break,
etc.) of an old, red elliptical galaxy.  However, the spectra often
have low S/N, and most (45/48) are limited by the 5600-9200{\AA}
wavelength range of Magellan/IMACS.  The 20 bands of high-S/N optical
photometry allow us to take a broader look at the optical SED.

We fit the optical photometry of each optically dull AGN with an
``r+q'' template that is a mix of a red galaxy component \citep[the
SDSS red galaxy composite from][]{eis01} and a blue AGN component
\citep[the SDSS quasar composite from][]{van01}.  The scale of each
component is an independent free parameter.  The two components of the
hybrid ``r+q'' template are well-motivated for two reasons: (a) from
the X-ray properties the objects must have an AGN, and (b) the optical
spectrum most closely resembles a red galaxy.  While both the host and
any underlying AGN will not be perfectly described by the ``r'' and
``q'' components of the template, we explore the minor systematic
deviations below.

We find the red galaxy and AGN components in the best-fitting template
by maximizing the Bayesian probability function, $P = \prod
\frac{1}{\sqrt{2\pi\sigma_m^2}} \exp{( \frac{-0.5 (m-m_t)^2}
{\sigma_m^2} )}$.  Here $m$ is the observed magnitude, $\sigma_m$ is
its error, and $m_t$ is the template magnitude computed by measuring
the template flux through the same wavelength response function as the
observed magnitude.  (The $\chi^2_0$ parameter is the logarithm of
this probability function, but we choose the Bayesian approach because
it maps out the probability distribution, not just the best-fit
values.)  In the fits for all objects the best-fit fractions of AGN
and red galaxy are tightly constrained: the 99\% confidence intervals
for the fit contain deviations of $<3\%$.

Systematic errors will dominate over the fitting errors, however,
because the ``r+q'' template is not likely to be a perfect fit to the
observed data.  First, the optically dull AGNs may have active or
recent star formation contributing to the blue emission, causing us to
overestimate the AGN emission.  The contribution from a young stellar
population (O/B-star) is likely to be minor, since the emission line
luminosities for the optically dull AGNs are very low (see \S 2.2).  A
moderate age (A-star) stellar population would not have strong
emission lines, but must also be a minor contributor at best because
none of the optically dull AGN spectra show a Balmer break.  We
estimate the effect of any blue star-forming component as $<20\%$,
since any higher contribution would lead to emission lines or a
recognizable Balmer break for even the lowest S/N optically dull AGNs.
In addition, the AGN template of \citet{van01} may not well describe
the optical shape of the true underlying AGN, since AGNs can be
heavily reddened \citep{hop04} or obscured \citep{eli08}, and even
unobscured quasars are known to exhibit a wide variety of optical
spectral shapes \citep{ric06}.  Still, \citet{van01} notes that the
Type 1 AGNs have a variation of only $\sigma<20\%$ from the mean SED
in their blue ($\lambda<4000$\AA) continua.  So we can assume that our
AGN fractions are valid, with the caveat that the true optical AGN
emission may differ by up to 20\%.  While the ``r+q'' template may not
recover the true optical AGN fraction of the optically dull AGNs, our
estimated AGN fraction is useful as rough estimate for studying the
host-subracted $X/O$ fraction.

\begin{figure}[ht]
\scalebox{1.2}
{\plotone{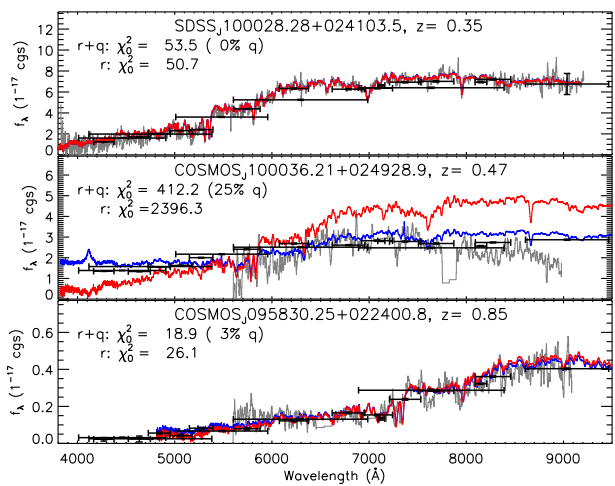}}
\caption{Three examples of our fits to the optical photometry of
optically dull AGNs.  In each panel the gray histogram shows the
spectrum, from the SDSS for the first object and from Magellan/IMACS
for the other two.  The black points with error bars show the measured
Subaru optical magnitudes.  The red and blue histograms show the
best-fit red galaxy (``r'') and quasar/galaxy hybrid (``r+q'')
templates, respectively.  In the upper left of each panel, we show the
reduced chi-square for the best fit, and the fraction of the hybrid
template from the quasar component is represented by ``(X\% q)''.
Note that the fitting comes from the photometry and does not use the
spectrum.  The three panels represent, from top to bottom: optically
dull AGNs with virtually no blue quasar emission, objects with
significant blue emission above the red galaxy host, and intermediate
objects.
\label{fig:photfits}}
\end{figure}

We show examples of our template fits in Figure \ref{fig:photfits}.
In each panel the black points show the photometry, with the x error
bar showing the band width and the y error showing the photometric
error.  The black histogram is the observed spectrum from
Magellan/IMACS or the SDSS, and the blue histogram is the best-fit
``r+q'' template.  Figure \ref{fig:photfits} additionally includes a
best-fit red-galaxy-only (``r'') template, shown in red, to illustrate
the improvement of including a blue AGN component in the template fit.
Reduced chi-square values, and the blue AGN contribution of the
``r+q'' template fit, are shown in the upper left of each panel.  Note
that the $\chi_0^2$ values are quite large because the optical
photometry has very small errors and the 14 narrow and intermediate
bands are sensitive to details which are not well-described by our
templates.  But while the fits do not perfectly describe the details
of the optically dull AGN SEDs, the templates are useful for studying
the shape of the SEDs and providing a rough estimate of the relative
blue AGN and red host components.

The top panel represents optically dull AGNs with virtually no quasar
contribution in the optical photometry, where the highest probability
``r+q'' template had zero quasar fraction.  Five of the 48 optically
dull AGNs had similar fits, with virtually no blue AGN emission.  The
upper limit on any blue quasar excess in these objects is typically
only 2\% blue AGN component.

The second panel of Figure \ref{fig:photfits} shows an example of an
optically dull AGN with a significant quasar component in the fit.  In
this example the best-fit ``r+q'' template is significantly better
than the best-fit ``r'' template, with a much lower reduced
chi-square.  The majority of the optically dull AGNs, 28/48 objects,
exhibited similar fits, with $\chi_0^2({\rm r}) \ge 2 \chi_0^2({\rm
r+q})$.  The blue AGN contribution in these objects is typically
15-35\%.

The third panel represents optically dull AGNs where the best-fit
``r+q'' template is only a slight improvement over the plain ``r'' red
galaxy template.  These AGNs have only a very weak excess of blue
emission, completely invisible in the observed spectrum and only
barely detected in the optical photometry.  Of the 48 optically dull
AGNs, 15 exhibit similar fits, with blue AGN contribution of about
3-7\%.

We can additionally compare the predicted line fluxes of the quasar
component in the best-fit template to the line flux limits in the
optical spectrum.  The H$\beta$ and {\OII} narrow emission lines
(shown in Figure \ref{fig:linelimits}) do not work well for this
comparison because we use a quasar template in our fit, and these
narrow lines are often weak or nonexistent in Type 1 AGN.  However the
{\OIII} ($\lambda$5007\AA) narrow emission line is typically strong in
both Type 1 and Type 2 AGN and so is useful for the comparison.  Only
30 optically dull AGNs have {\OIII} in their observed wavelength
range, and all of these are upper limits only.  Most (19/30) of these
AGNs have predicted {\OIII} fluxes from the best-fit model which lie
below the upper limit on {\OIII} flux from the spectrum.  Since these
AGNs have low predicted line fluxes and the measured {\OIII} fluxes
are only limits, it is not a strong constraint, but it does suggest
that these 19 optically dull AGNs could be diluted Type 1 or Type 2
AGNs.

\begin{figure}[ht]
\scalebox{1.2}
{\plotone{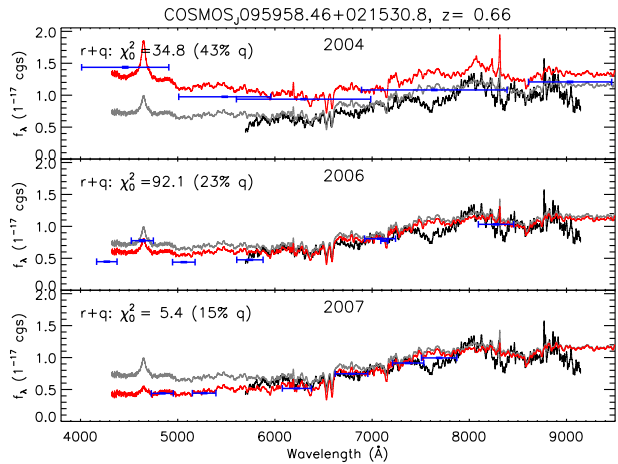}}
\caption{An optically dull AGN which exhibits variability in its
optical photometry.  In all three panels, the black histogram shows
the observed Magellan/IMACS spectrum (taken in 2006), and the gray
histogram shows the best-fit ``r+q'' hybrid template to the 20-band
photometry from all years.  Each panel represents a different year of
observations: 2004 at the top, 2006 in the middle, and 2007 at the
bottom.  The blue points with error bars show the photometry for that
year and the red histogram is the best-fit ``r+q'' template for that
year's data only.  The optically dull AGN has a strongly decreasing
blue emission component, starting as nearly half of the total emission
in 2004 and dropping to less than one-sixth in 2007.  Four optically
dull AGNs show significant variability.
\label{fig:varfits}}
\end{figure}

The optical photometry also reveals significant variability in four
optically dull AGNs.  When comparing the observations from 2004, 2006,
and 2007 (see Table \ref{tbl:photometry}), these four AGNs exhibited
changes in flux 5$\sigma$ beyond the photometry errors.  We show an
example of a variable optically dull AGN in Figure \ref{fig:varfits}.
Magnitudes from each of 2004, 2006, and 2007 are shown in each panel
in blue, with the corresponding template fit shown in red in each
panel.  The template fit to all 20 bands of optical photometry (from
all years) is shown in gray, along with the Magellan/IMACS spectrum in
black, for comparison in each panel.  The optically dull AGN decreases
in flux from 2004 to 2007, but almost all of this change is in the
blue emission.  In the template fit, the red galaxy component remains
nearly the same in each year while the AGN component decreases from
43\% to 15\% contribution.

Old red galaxies do not change in flux over different years of
observations.  Type 1 AGNs, however, can vary by as much as factors of
a few on year timescales \citep[e.g.][]{mor08, kel09}.  The four
variable optically dull AGNs must then have a weak AGN causing the
variability.  The source of the variability must be $\le$1 light-year
in size, making obscuration or reddening extremely unlikely.  The
variable optically dull AGNs are instead likely to be diluted
``normal'' AGNs.  Indeed, close inspection of Figure \ref{fig:varfits}
shows that the optical spectrum may have a weak H$\beta$ broad
emission line, although it is difficult to positively identify the
line because of low S/N in that part of the spectrum.  (\S 2.2,
however, showed that for this and other optically dull AGNs, the
narrow {\OII} and H$\beta$ lines are not hidden by low S/N, but are
instead very weak compared to those of Type 2 AGNs.)  So while
optically dull AGNs do not have strong emission lines, the four
variable objects in particular show evidence for a diluted (not
obscured) AGN.  These objects are likely to be normal, unobscured Type
1 AGNs diluted by extranuclear light (as we explore in \S 3.5).

\subsection{X-ray to Optical Ratio}

The defining characteristic of optically dull AGNs is that they are
bright in X-rays while their optical spectra have no sign of emission
lines.  But while optically dull AGNs lack the emission line signature
of an AGN, \S3.1 showed that they do have excess blue emission which
might be attributed to a diluted AGN.  But are optically dull AGNs
simply diluted by a bright host, or is their optical emission actually
depressed when compared to their bright X-rays?

\begin{figure}[ht]
\scalebox{1.2}
{\plotone{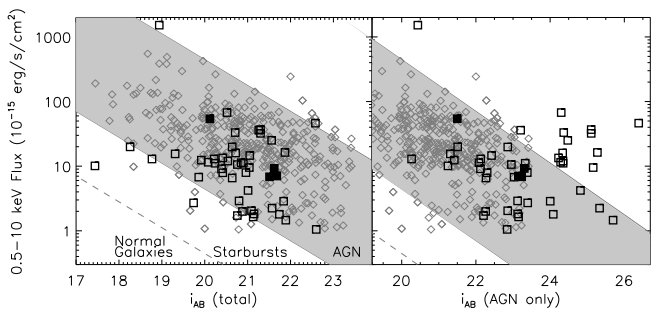}}
\caption{The 0.5-2 keV X-ray flux and $i_{\rm AB}$ optical magnitude
for the optically dull AGNs, shown by black squares.  Filled squares
indicate the four optically dull AGNs with significant variability.
Also shown by gray diamonds are the Type 1 and Type 2 X-ray AGNs of
\citet{tru09a}.  In each panel the gray shaded area is the traditional
AGN locus \citep{mac88}, and the $f_X/f_O$ regions of normal galaxies
and starbursts are additionally indicated in the lower left.  The left
panel uses the full $i_{\rm AB}$ flux from both host and AGN, while
the right panel uses only the AGN flux from our photometric fitting.
Even with the host contribution removed, 33 of the 48 optically dull
AGNs do not lie outside the X-ray AGN $f_X/f_O$ locus.
\label{fig:xraylocus}}
\end{figure}

We present the ratio between the X-ray and optical flux in Figure
\ref{fig:xraylocus}, where $\log{f_X/f_O}=\log(f_X) + i_{\rm AB}/2.5 +
5.352$.  The optically dull AGNs are shown as squares, and the four
variable objects are indicated by filled squares.  For comparison the
Type 1 (broad line) and Type 2 (narrow line) X-ray AGNs of
\citet{tru09a} are shown in gray.  At left, we use the total $f_O$ for
the optically dull AGNs, and all but 2 of the optically dull AGNs have
$f_X/f_O$ values consistent with typical AGNs.  At right, the $i_{\rm
AB}$ magnitude includes only the AGN fraction as determined in \S 3.1.
It is important to note that X-ray K-corrections will cause
Compton-thick AGNs at higher redshifts to have higher $f_X/f_O$
ratios, \citep[e.g.,][]{com03}, although this effect should be minimal
in our sample because very few of the optically dull AGNs are
Compton-thick (see \S 3.4) and all have $z<1$.

Even after subtracting out the host component, 33/48 optically dull
AGNs have $f_X/f_O$ values consistent with typical AGNs.  These
optically dull AGNs might be normal AGNs diluted by their hosts.
However, we note that host dilution should push objects to the left in
Figure \ref{fig:xraylocus}, so host dilution may be unlikely for AGNs
with $f_X/f_O \sim 1$ and 33/48 may be an upper limit on the true
fraction of optically dull AGNs diluted by their hosts.  The 15 AGNs
with $f_X/f_O>1$ present the most interesting case, since host
dilution is impossible and some physical effect must depress their
optical emission while they remain X-ray bright.

\subsection{Infrared Color: Dust Properties}

Bright AGNs are well-known to have redder {\em Spitzer}/IRAC colors
than normal galaxies \citep{lacy04, stern05} as a result of strong
mid-IR power-law continua \citep{saj05, don07}.  The IRAC emission is
generally associated with the hot, dusty ``torus'' or outer accretion
disk of AGNs.  Since optically dull AGNs are optically fainter than
normal AGNs, they might also have different mid-IR properties, with
the power-law continuum either diluted or absent.

\begin{figure}[ht]
\scalebox{1.2}
{\plotone{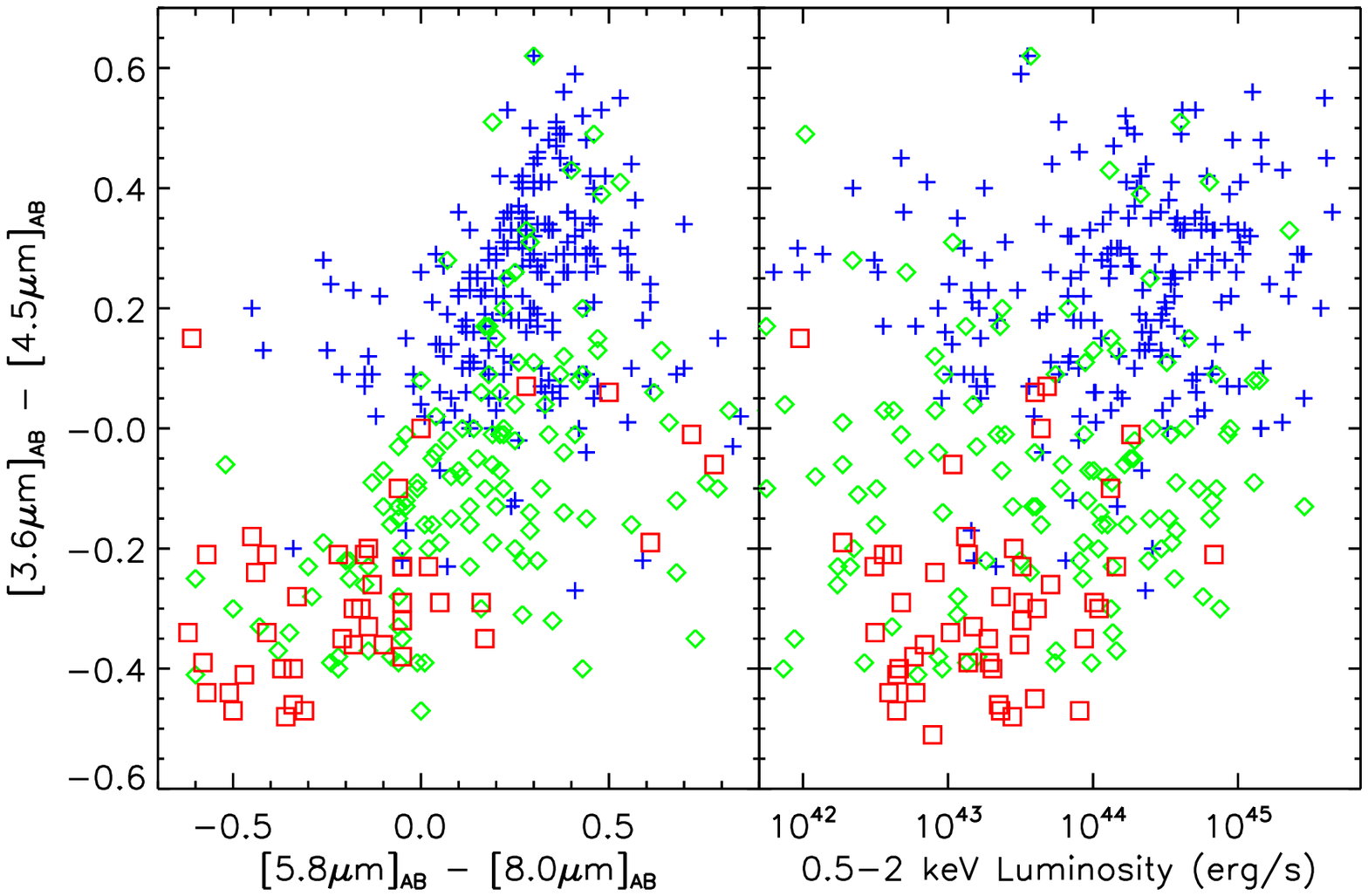}}
\caption{At left, {\em Spitzer}/IRAC colors are shown for the
optically dull AGNs (red squares), along with Type 1 (broad-line) AGNs
(blue crosses) and Type 2 (narrow-line) AGNs (green diamonds) for
comparison \citep[from][]{tru09a}.  Bright Type 1 AGNs are known to
have redder IRAC colors due to their strong red continua, indicative
of hot dust \citep{lacy04, stern05, saj05}.  Most of the optically
dull AGNs have IRAC colors consistent with normal galaxies,
significantly bluer than Type 1 and Type 2 AGNs.  At right, we show
the $[3.6{\mu}m]_{\rm AB} - [4.5{\mu}m]_{\rm AB}$ color with the X-ray
luminosity.  Optically dull AGNs have similar X-ray luminosities to
Type 1 and Type 2 AGNs, even as their IRAC colors are markedly
different.
\label{fig:irac}}
\end{figure}

The IRAC colors are shown in Figure \ref{fig:irac}.  The optically
dull AGNs are marked with red squares, while Type 1 and Type 2 AGNs
from \citet{tru09a} are shown as blue crosses and green diamonds,
respectively.  Type 1 AGNs generally have the reddest IRAC colors,
followed by Type 2 AGNs, while most optically dull AGNs have IRAC
colors consistent with normal galaxies.  The $[3.6{\mu}m]_{\rm AB} -
[4.5{\mu}m]_{\rm AB}$ color does a particularly good job of separating
the various AGN types.  We additionally show the $[3.6{\mu}m]_{\rm AB}
- [4.5{\mu}m]_{\rm AB}$ color with soft X-ray luminosity in the right
panel of Figure \ref{fig:irac}.  \citet{don07} suggested that the
mid-IR power-law continuum disappears at low X-ray luminosities, but
this does not appear to be the case for our optically dull AGNs.
While many Type 1 AGNs are more X-ray luminous than the optically dull
AGNs, many have similar luminosities, and there is no apparent
correlation between IRAC color and X-ray luminosity in Figure
\ref{fig:irac}.  The optically dull AGNs have IRAC colors consistent
with normal galaxies even though they are as X-ray luminous as some
Type 1 AGNs.

\subsection{X-ray Column Density}

Several authors have suggested that the optical emission of optically
dull AGNs is obscured, either by material near the central engine
\citep{com02, civ07} or by gas and dust in the host galaxy
\citep{rig06}.  But if the optical emission is obscured, then the
X-ray emission would probably be obscured as well (so long as the
obscuring material for X-ray and optical emission is cospatial).  For
the 28 optically dull AGNs with $>$50 full band counts in their XMM or
Chandra \citep{elv09, lan09} observations, we estimate $N_H$ from
X-ray spectral analysis.  We fit each X-ray spectrum as an
intrinsically absorbed power-law with Galactic absorption ($N_{H,{\rm
gal}} = 2.6 \times 10^{20}$~cm$^2$ in the direction of the COSMOS
field), with the power-law slope and $N_H$ as free parameters.  The
best-fit $N_H$ value and its 2$\sigma$ error are found using the
\citet{cash} statistic.  For the remaining 20 optically dull AGNs, we
estimate a less accurate $N_H$ from their hardness ratio,
$HR=(H-S)/(H+S)$, following the relation between $N_H$ and $HR$ from
\citet{mai07}.  Here $H$ is the counts in the hard 2-4.5 keV XMM band
and $S$ is the counts in the soft 0.5-2 keV XMM band.

\begin{figure}[ht]
\scalebox{1.2}
{\plotone{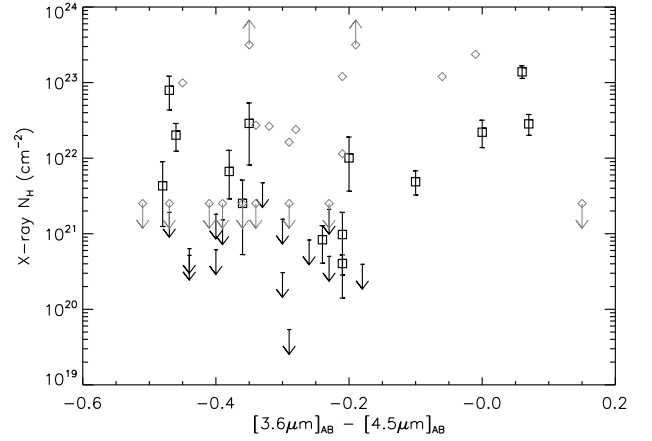}}
\caption{The X-ray column density $N_H$ with the $[3.6{\mu}m]_{\rm
AB}-[4.5{\mu}m]_{\rm AB}$ {\em Spitzer}/IRAC color for the optically
dull AGNs.  For the 28 optically dull AGNs shown as black squares or
upper limits, the X-ray or Chandra observation had more than 50 counts
and a reliable $N_H$ was calculated directly from the X-ray spectrum.
The remaining 20 optically dull AGNs have less reliable column
densities inferred from their hardness ratio and are shown in gray.
Optically dull are not highly X-ray obscured, and instead have similar
$N_H$ to Type 2 AGNs in COSMOS \citep{mai07}.
\label{fig:nh}}
\end{figure}

Figure \ref{fig:nh} shows the column density $N_H$ with the
$[3.6{\mu}m]_{\rm AB} - [4.5{\mu}m]_{\rm AB}$ color.  Black squares
and upper limits show those optically dull AGNs with over 50 counts in
their XMM or Chandra observation, while gray diamonds show those
objects with $N_H$ estimates from the hardnress ratio only.  Most (31)
optically dull are relatively unobscured in their X-rays, with
$N_H<10^{22}$~cm$^{-2}$, and at most only 2-3 are Compton-thick
($N_H>10^{24}$~cm$^{-2}$).  The X-ray column densities are similar to
those of Type 1 and Type 2 AGNs in COSMOS \citep{mai07}, so there is
no X-ray evidence for additional obscuration in optically dull AGNs.

\subsection{Host Galaxy Properties}

\begin{figure*}[ht]
\scalebox{1.15}
{\plotone{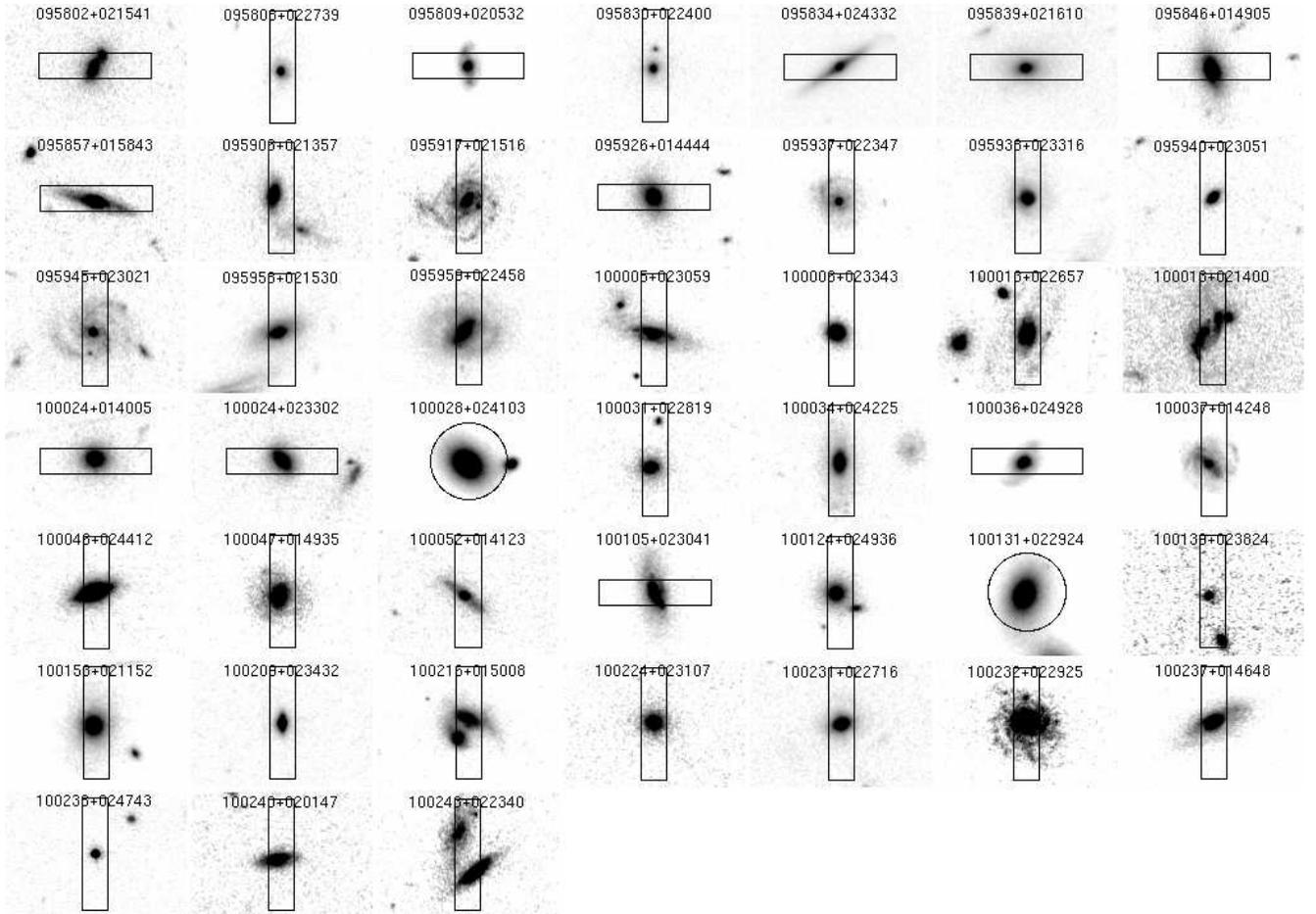}}
\caption{HST/ACS images of 45 optically dull AGNs.  In each 6\farcs4
by 5\farcs6 image the spectroscopic slit or fiber is overlayed in
black (IMACS spectra use $1{\farcs}0 \times 5{\farcs}4$ slits, while
SDSS spectra use $3{\farcs}0$-diameter fibers).  The host galaxies of
the optically dull AGNs have a large range of morphological types and
orientations.  In several images, it is clear that a nearby companion
galaxy or the orientation of host causes significant extranuclear
light to fall within the spectroscopic slit.
\label{fig:hosts}}
\end{figure*}

The HST/ACS imaging in COSMOS allows for detailed studies of the host
galaxies of the optically dull AGNs.  We show postage stamps of the 46
objects with HST/ACS coverage in Figure \ref{fig:hosts}.  Immediately
it is evident that the optically dull AGNs reside in a wide variety of
hosts \citep[in contrast with][]{rig06}, despite the fact that they
have spectra consistent with old, red elliptical galaxies.  A few
hosts appear as isolated spheroids or ellipticals, while others have
clumpy and dusty disks, and some are disturbed.  Type 1 and Type 2
AGNs have similarly been shown to exist in a wide range of host galaxy
morphologies \citep{jah04, san04, gab09}.  Marking the spectroscopic
aperture ($1{\farcs}0 \times 5{\farcs}4$ IMACS slit or
$3{\farcs}0$-diameter SDSS fiber) over each of the images, however,
reveals a common thread: several of the optically dull AGNs appear to
be have significant extranuclear light within the aperture.  Both of
the optically dull AGNs with SDSS spectroscopy in Figure
\ref{fig:hosts} have bright elliptical hosts filling the fiber
aperture.  At least 8 objects with IMACS spectroscopy have a nearby
companion falling in the slit, while the hosts of at least 8 others
appear to have a bar or disk oriented along the slit.  In all of these
cases, the AGN optical emission is likely to be diluted by the
continua of one or more normal galaxies.  This scenario can explain
the optically dull AGNs with normal $f_X/f_O$ ratios, since the
extranuclear host galaxy light would increase the total optical
brightness.

\begin{figure}[ht]
\scalebox{1.2}
{\plotone{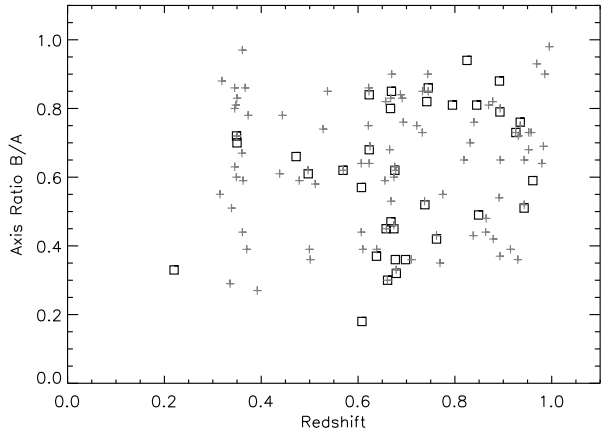}}
\caption{The axis ratio $b/a$ with redshift for 37 optically dull AGNs
(black squares), with 72 Type 2 AGNs from \citet{gab09} shown by gray
crosses.  High values of $b/a$ indicate face-on or spheroidal systems,
while edge-on disks have low $b/a$.  The axis ratios of optically dull
AGNs are quite similar to those of Type 2 AGNs, with no preference for
edge-on or face-on systems.
\label{fig:axisratio}}
\end{figure}

\citet{rig06} used HST/ACS images in the Chandra Deep Field South
(CDF-S) to show that optically dull AGNs have preferentially edge-on
hosts compared to other AGNs, further suggesting that optically dull
AGNs are optically obscured by extranuclear dust in their host
galaxies.  To test this hypothesis, we present axis ratios with
redshift for 37 optically dull AGNs in Figure \ref{fig:axisratio},
along with a sample of 93 Type 2 AGN from \citet{gab09} in gray.  No
Type 1 AGN are shown because the point sources are too bright for
accurate host galaxy decompositions: in \citet{gab09}, $2/3$ of Type 1
AGN hosts had unphysical best-fit parameters.

We measure $b/a$, the ratio of minor to major axis, using the publicly
available galaxy fitting software GALFIT \citep{pen02} and following
the procedures of \citet{gab09}.  Sources in the AGN vicinity
($<35$h$^{-1}$kpc projected on the sky) are identified using Source
Extractor \citep{ber96} and are either masked out of the image or
simultaneously fit by GALFIT (if they are too close to the AGN for
masking).  We use the Source Extractor outputs based on isophotal
profiles to generate initial guesses for the magnitude and shape
parameters of the galaxy images in GALFIT.  We fit each AGN image two
separate times to explore different light distribution models.  In one
fit, we model the galaxy as a single S\'{e}rsic profile, and another
uses a S\'{e}rsic profile plus a point source representing a nuclear
point source (although over half of the optically dull AGNs have no
strong nuclear point source; see below).  \citet{gab09} explored fits
with additional components (e.g. a disk $+$ S\'{e}rsic profile), but
found that such fits typically give unphysical results because they
are unstable for the S/N of the HST/ACS images.  By constraining the
fits in radius, magnitude, and shape, we prevent GALFIT from exploring
wildly unphysical parameter space, but we flag as unacceptable any
fits which run into the boundaries or yield strikingly unphysical
results \citep[for more details, see][]{gab09}.

The GALFIT analysis yields good fits for 37 of the optically dull
AGNs.  Most (21) of the optically dull AGNs are best fit with a single
S\'{e}rsic component and no nuclear point source.  Of the remaining
16 with two-component fits, 13 have only marginal contributions from a
point source, and fitting these 13 AGNs with S\'{e}rsic-only
components does not affect their $b/a$ values.  Only 3 optically dull
AGNs have significant nuclear point sources which would bias their
$b/a$ measurements to high (less elongated) values if not included in
the fit.  These nuclear point source contributions are consistent
with the tempate fitting in \S 3.1, which showed that the optically
dull AGNs have blue AGN contributions of 35\% or less.  The GALFIT
axis ratios correlate strongly with those measured using Source
Extractor, with a mean absolute difference of 0.11 in $b/a$.  This
suggests that our axis ratios are robust.

The optically dull AGNs and Type 2 AGNs in COSMOS have nearly
identical ranges of axis ratio, with the optically dull AGN mean
$b/a=0.56 \pm 0.20$ and the Type 2 AGN mean $b/a=0.56 \pm 0.18$.
While our optically dull AGNs have consistent axis ratios to those of
\citet{rig06}, our Type 2 AGNs do not show the face-on preference that
\citet{rig06} claim for their ``optically active'' AGN sample.  Part
of this difference comes from the differences in sample definitions:
the 6 ``optically active'' AGNs of \citet{rig06} include 4 broad-line
Type 1 AGNs, while we compare to only Type 2 AGNs.  \citet{rig06}
showed that optically dull AGNs are quite different from Type 1 AGNs,
while our Figure \ref{fig:axisratio} shows that optically dull AGNs
have similar hosts to Type 2 AGNs.  In \citet{gab09}, it was shown
that morphological fits to Type 1 AGN hosts suffer from many
systematic errors.  In particular, a Type 1 host could have an
incorrectly high $b/a$ value, since even a slightly incorrect point
source removal would leave a symmetric halo and a corresponding round
residual.  In any case, the fact that Type 2 and optically dull AGNs
have similar axis ratios indicates that edge-on hosts are not causing
the lack of narrow emission lines in optically dull AGNs.

\section{Discussion}

Combining the optical, X-ray, and infrared data, we have shown that
optically dull AGNs exhibit the following properties:

\begin{enumerate}
  \item Nearly all (43/48) optically dull AGNs have significantly more
    blue emission than a typical red galaxy.
  \item A few (4/48) optically dull AGNs show variability on year
    timescales, especially in their blue emission.
  \item Even when counting only the blue AGN component, $\sim$70\%
    (33/48) of optically dull AGNs have $f_X/f_O$ ratios like typical
    Type 1 and 2 AGNs.
  \item Optically dull AGNs lack the mid-IR power-law signature of
    Type 1 and Type 2 AGNs, instead exhibiting cool IRAC colors like
    normal galaxies.
  \item The X-ray column densities of optically dull AGNs are similar
    to those of Type 1 and Type 2 AGN, with no evidence for more
    absorption.
  \item Optically dull AGNs reside in a wide morphological variety of
    host galaxies, including isolated ellipticals, dusty spirals, and
    disturbed and potentially merging systems.
  \item At least 18/45 optically dull AGNs with HST/ACS imaging are
    diluted by extranuclear light in the spectroscopic aperture,
    either by a nearby companion galaxy or host galaxy light.
  \item The hosts of optically dull AGNs are not preferentially
    edge-on compared to Type 2 AGNs, so edge-on host galaxy
    obscuration cannot explain the lack of narrow emission lines.
\end{enumerate}

While several authors \citep{com02, rig06, civ07} have suggested that
optically dull AGNs are optically obscured, we find no evidence for
Compton-thick or hot toroidal obscuration.  While we can't rule out
weak obscuration \citep[as proposed by][]{civ07}, the $N_H$ values for
optically dull AGN are fully consistent with those of Type 2 AGNs
\citep{mai07}, and Type 2 AGNs have emission lines while optically
dull AGNs do not.  Instead, our data support a framework where
$\sim$70\% (33/48) of optically dull AGNs are normal AGNs diluted by
extranuclear galaxy light.  The remainder of optically dull AGNs are
not diluted or obscured, but have different emission properties for
physical reasons: possibly because of a radiatively inefficient
accretion flow.

\subsection{The Case for Dilution}

At the redshifts of the optically dull AGNs in the sample, our
spectroscopic slit generally includes nearly all of the host galaxy,
and occasionally even includes a nearby companion.  This is especially
evident in Figure \ref{fig:hosts}, where at least 10 host galaxies
contaminate the spectroscopic aperture and at least 8 others have a
companion galaxy in the slit.  One can imagine that many ``optically
normal'' local Seyfert AGNs would appear ``optically dull'' if
observed with spectroscopic apertures including extranuclear galaxy
emission.  Indeed, \citet{mor02} obtained integrated spectra for 18
local Seyfert 2 galaxies, and found that 11 ($\sim$60\%) of them would
appear optically dull when observed in a $5\arcsec \times 1\arcsec$
spectroscopic slit at $z \gtrsim 0.5$.  Many of our optically dull
AGNs may then be analogs to local Seyfert 2 AGNs.  Dilution provides
the simplest explanation for the four variable optically dull AGNs,
all of which have a clear blue component in their optical photometry
and $(f_X/f_O)<1$ for the AGN fraction of the template fit.  Dilution
by a host galaxy might explain all 33/48 (70\%) of the optically dull
AGN with $f_X/f_O$ ratios consistent with Type 1 and Type 2 AGN (that
is, $\log(f_X/f_O)<1$).  While only 18 show obvious evidence for
extranuclear galaxy light in the slit, the other $\log(f_X/f_O)<1$
objects might be weak AGN with the emission lines diluted by a bright
host.  AGN activity is typically correlated with host luminosity
\citep[e.g.,][]{hic09, sil09}, but there is a large scatter in the
relation.  Under the dilution hypothesis, some optically dull AGNs may
represent the weak AGN / bright host tail of the relation.

However, dilution cannot explain all optically dull AGNs.  Locally,
10-20\% of local AGNs are undiluted and remain optically dull
\citep{laf02, hor05}.  And in COSMOS, 15 optically dull AGNs are
optically under-luminous compared to their X-ray emission, with
$\log(f_X/f_O)>1$.  Dilution by a host galaxy, on the other hand,
would cause AGNs to become more optically luminous compared to their
X-ray emission.  Indeed, optically dull AGNs with $\log(f_X/f_O) \sim
1$ may also not fit the dilution paradigm, since presumably the
additional host light would drive the optical flux of ``normal'' AGNs
well below this cutoff.  This suggests that 15/48 ($\sim$30\%) is a
lower limit for the optically dull AGNs not explained by dilution.

\subsection{The Case for Radiatively Inefficient Accretion}

The optically dull AGNs in COSMOS do not show signs of strong
obscuration, with X-ray column densities similar to Type 2 AGNs and
blue IRAC colors.  Their host galaxies are not preferentially edge-on
compared to the hosts of Type 2 AGNs, suggesting that obscuration by
the host is not the cause of their missing narrow emission lines.
With no evidence for obscuration, the undiluted optically dull AGNs
must be intrinsically weak in their optical emission.  AGNs with low
accretion rates are expected to be optically underluminous, with very
weak or missing emission lines, in just this fashion.  In the next
section we investigate the properties of the 15/48 (30\%) optically
dull AGNs which are not explained by obscuration or dilution to see if
they fit the properties expected for low accretion rate AGNs.

\section{Accretion Properties}

Observations have shown that broad lines tend to disappear from AGNs
below accretion rates of $L/L_{Edd} \sim 0.01$ \citep{kol06,tru09b}.
Type 1 AGNs are likely to decay into ``naked'' Type 2 AGNs
\citep{tra03, bia08} which have no evidence for obscuration.  The
theoretical interpretation \citep{nen08, hop09, eli09} is that the
broad line region decays as a natural effect of a shrinking accretion
disk below $L/L_{Edd} \sim 0.01$, even as the X-ray emission remains
bright.  The undiluted optically dull AGNs may then be an extension of
these ideas, with lower accretion rates driving an altered accretion
disk.  In the paradigm most suited to explaining optically dull AGNs,
the accretion disk is optically thick as normal at higher radii from
the black hole, but becomes optically thin below some transition
radius \citep{yua04}.  Thus the hot optical and UV continuum becomes
cooler and redder, and the ionizing continuum becomes much weaker.
Without an ionizing continuum, neither the broad nor the narrow line
regions are excited, and the spectrum lacks the emission line
signature of an AGN.  \citet{ho99} noted that several local
low-luminosity AGNs exhibit this behavior, with a generally redder
optical/UV continuum and a lack of the strong UV ``big blue bump''
found in luminous AGNs.  Unfortunately, we cannot measure the
accretion rate $L/L_{Edd}$ for these optically dull AGN because we
cannot measure the black hole mass: they lack lines for using the
scaling relations and they are too distant for dynamical estimates.
Future work may leverage the $M_{BH}-M_{bulge}$ relation to estimate
$L/L_{Edd}$, but that is beyond the scope of this work.  Instead we
will study other properties to see if these optically dull AGN are
consistent with predictions of RIAF models.

\begin{figure}[ht]
\scalebox{1.2}
{\plotone{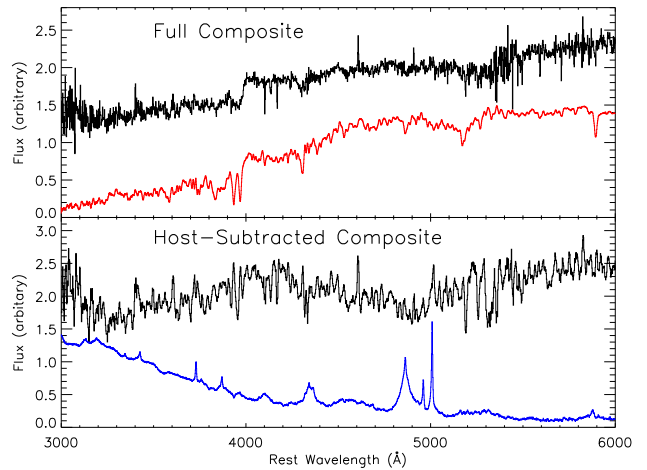}}
\caption{The composite spectrum of the 15 optically dull AGNs with
$\log(f_X/f_O)>1$ (those likely to have RIAFs).  The top panel shows
the full composite, with a red galaxy composite \citep{eis01} shown in
red for a comparison.  The bottom panel shows the composite of the
host-subtracted spectra, with a Type 1 AGN composite \citep{van01}
shown in blue.  The full composite is bluer than a simple red galaxy,
and neither composite shows evidence for narrow or broad emission
lines.  The host-subtracted composite is much redder than a typical
Type 1 AGN.  This suggests that these optically dull AGNs may have
RIAFs, where the optical accretion disk is truncated at lower radii,
causing less UV emission and a weaker ionizing continuum.
\label{fig:adafspec}}
\end{figure}

Figure \ref{fig:adafspec} shows the 3000-6000{\AA} optical/UV
composite spectrum from the 15 optically dull AGNs likely to have
RIAFs.  The top panel shows the full composite, while the bottom panel
uses only the host-subtracted spectra to compute an AGN-only
composite.  Each composite spectrum was computed by taking a
S/N-weighted mean of the spectra.  (The absorption features of the
full composite spectrum are artificially broadened by minor redshift
errors in some of the optically dull AGN.)  Note that the AGN-only
composite is not simply the full composite minus a mean host
component, but was computed from the individual host-subtracted
spectra, using the best-fit ``r+q'' template from \S 3.1.  For
comparison, Figure \ref{fig:adafspec} also shows SDSS composites of a
red galaxy \citep{eis01} and a Type 1 AGN \citep{van01}.  Neither
composite has broad or narrow emission lines, despite having higher
S/N than the individual optically dull AGN spectra.  While the full
composite is bluer than a typical red galaxy, the AGN-only composite
is much redder than a typical Type 1 AGN.  The optical/UV instead
supports a RIAF model with a truncated accretion disk and less hot UV
emission.

\begin{figure}[ht]
\scalebox{1.2}
{\plotone{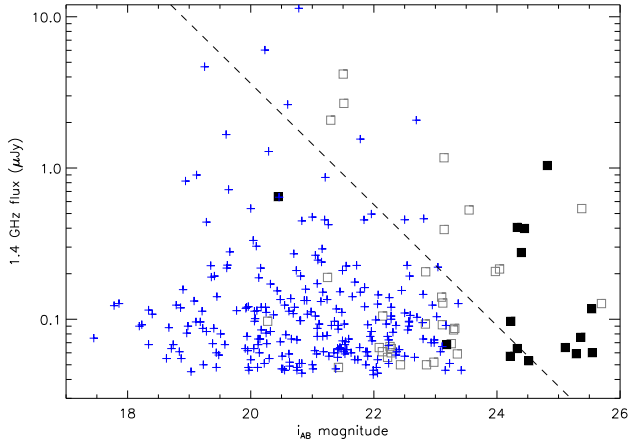}}
\caption{The 1.4 GHz radio fluxes and $i_{\rm AB}$ magnitudes for the
optically dull AGNs (black squares), along with Type 1 AGNs (blue
crosses) from \citet{tru09a}.  The optically dull AGN with
$\log(f_X/f_O)>1$ (those likely to have RIAFs) are shown by the filled
squares.  Optical magnitudes for the optically dull AGNs include only
the AGN component, as estimated from the best-fit ``r+q'' template in
\S 3.1.  While the Type 1 AGNs may include some host galaxy light, it
is likely minor relative to the AGN contribution.  The dashed line
represents $f_{1.4~Ghz}/f_i=10$.  Optically dull AGNs, especially
those with $\log(f_X/f_O)>1$, are more likely to be more radio-loud
than Type 1 AGNs.
\label{fig:radio}}
\end{figure}

Another important prediction for low accretion rate AGNs is that the
dominant outflow mode switches from disk winds to radio jets
\citep{ho02, eli06}.  The radio synchrotron emission provides the
dominant source of cooling and angular momentum transport for RIAF
accretion states as the optically thin inner accretion disk no longer
drives strong disk winds.  Since the RIAF truncated accretion disk is
naturally under-luminous in the optical, low accretion rate AGN should
be both radio-bright and optically dim: in other words, their
radio-loudness $f_R/f_O$ will be large.  The radio properties for the
optically dull AGN are shown in Figure \ref{fig:radio}, along a
comparison sample of Type 1 AGNs from \citet{tru09a}.  Radio data are
available from Very Large Array (VLA) observations in COSMOS
\citep{sch07}, and all optically dull AGN are detected at 1.4 GHz at
the $>5\sigma$ level.  The optically dull AGNs in Figure
\ref{fig:radio} include only the AGN component in the optical
emission.  The Type 1 AGNs in the figure may also include optical
light from the host galaxy, but any host light is likely to be minor
compared to the Type 1 AGN optical emission.  Nearly all of the
optically dull AGN with $\log(f_X/f_O)>1$ are radio-loud, with
$f_{1.4~Ghz}/f_i>10$, and they are more likely to be radio-loud than
Type 1 AGNs.  Their radio-loudness suggests that these optically dull
AGNs are RIAFs with a strong radio jet.

While a full fit of the optically dull AGN SEDs to low accretion rate
models is beyond the scope of this work, we have shown that the
RIAF-candidate optically dull AGN have optical/UV and radio properties
that differ from those of bright Type 1 and 2 AGNs.  We note again
that $\log(f_X/f_O)>1$ is a conservatively low limit for no host
galaxy dilution, and so the fraction of optically dull AGNs that are
RIAFs is likely to be greater than $\sim$30\%.  Indeed, Figure
\ref{fig:radio} shows that several optically dull AGNs with
$\log(f_X/f_O) \le 1$ are also radio-loud, lending one piece of
evidence that even more than 30\% of optically dull AGNs are in a
weakly accreting RIAF state.

\section{Summary}

We have presented 48 optically dull AGNs from COSMOS, all of which
lack optical emission lines while exhibiting the X-ray brightness
typical of an AGNs.  Their IR and X-ray emission show no evidence for
obscuration in excess of that in Type 1 and 2 AGNs, and their host
galaxies are not preferentially edge-on when compared to Type 2 AGNs.
We instead propose a framework where up to 70\% of optically dull AGNs
are diluted by their host galaxies or by nearby companions.  The
remaining 30\% cannot be explained by dilution, and instead have
optical/UV and radio properties which are best described by a RIAF
state.

\acknowledgements

JRT acknowledges support from NSF ADP grant NNX08AJ28G, an ARCS
fellowship, and a NSF-EAPSI/JSPS summer research fellowship.  BCK
acknowledges support from NASA through Hubble Fellowship grant
\#HF-01220.01 awarded by the Space Telescope Science Institute, which
is operated by the Association of Universities for Research in
Astronomy, Inc., for NASA, under contract NAS 5-26555.  We thank
J. Rigby for comments on the axis ratios of AGN host galaxies, and
A. Diamond-Stanic for help regarding line luminosity measurements.  We
additionally thank the anonymous referee for several suggestions which
improved the quality of this paper.

\begin{footnotesize}
\begin{deluxetable}{lccccc|cccc}
\tablecolumns{10}
\tablecaption{Optically Dull AGN Properties\label{tbl:odagn}}
\tablehead{
  \colhead{Object Name} & 
  \colhead{z} & 
  \colhead{S/N} & 
  \colhead{$i_{\rm AB}$} & 
  \colhead{$\log(L_X)$} & 
  \colhead{$X/O$} & 
  \colhead{$f_{\rm AGN}$} & 
  \colhead{$X/O$(AGN)} & 
  \colhead{$HR$\tablenotemark{a}} & 
  \colhead{$b/a$\tablenotemark{b}} }
\startdata
COSMOS J095802.10+021541.0 & 0.94 &  3.75 & 21.01 & 43.29 & -0.6 & 0.00 &  1.1 & -1.00 & -1.00 \\
COSMOS J095808.98+022739.9\tablenotemark{c} & 0.85 &  8.16 & 21.62 & 43.51 & -0.0 & 0.21 &  0.6 & -0.27 &  0.73 \\
COSMOS J095809.45+020532.4 & 0.61 &  7.73 & 20.80 & 42.65 & -0.9 & 0.12 &  0.1 & -1.00 &  0.57 \\
COSMOS J095820.57+023330.1\tablenotemark{c} & 0.96 &  1.18 & 21.51 & 43.52 & -0.2 & 0.20 &  0.5 & -0.42 &  0.60 \\
COSMOS J095830.25+022400.8 & 0.85 &  5.51 & 21.57 & 42.89 & -0.7 & 0.03 &  0.8 & -1.00 & -1.00 \\
COSMOS J095834.23+024332.5 & 0.39 & 14.33 & 19.86 & 42.56 & -0.9 & 0.05 &  0.4 &  0.20 &  0.27 \\
COSMOS J095839.01+021610.6 & 0.68 & 10.42 & 20.08 & 43.36 & -0.6 & 0.02 &  1.1 & -0.35 &  0.62 \\
COSMOS J095846.02+014905.6 & 0.74 &  5.75 & 20.24 & 43.44 & -0.5 & 0.06 &  0.7 & -0.10 &  0.52 \\
  SDSS J095849.02+013219.8 & 0.36 &  3.55 & 18.94 & 44.83 &  1.1 & 0.25 &  1.7 & -0.64 & -1.00 \\
COSMOS J095857.20+015843.7 & 0.52 &  6.02 & 21.30 & 43.60 &  0.4 & 0.03 &  1.9 &  0.66 &  0.34 \\
COSMOS J095906.97+021357.8 & 0.76 &  5.11 & 21.30 & 43.94 &  0.4 & 0.02 &  2.0 &  0.02 &  0.42 \\
COSMOS J095917.26+021516.9 & 0.94 &  7.88 & 20.85 & 43.71 & -0.2 & 0.32 &  0.2 & -0.51 &  0.76 \\
COSMOS J095926.01+014444.3 & 0.67 &  6.03 & 20.72 & 43.49 & -0.2 & 0.04 &  1.3 & -0.56 &  0.83 \\
COSMOS J095937.42+022347.4 & 0.74 &  5.90 & 21.15 & 42.66 & -0.9 & 0.16 & -0.1 & -1.00 &  0.82 \\
COSMOS J095938.56+023316.8 & 0.75 & 27.31 & 19.92 & 43.51 & -0.6 & 0.25 &  0.0 & -0.37 & -1.00 \\
COSMOS J095940.86+023051.2 & 0.70 &  7.16 & 21.74 & 42.59 & -0.7 & 0.12 &  0.2 & -1.00 &  0.34 \\
COSMOS J095945.21+023021.5 & 0.89 &  5.56 & 20.76 & 42.84 & -1.1 & 0.26 & -0.5 & -1.00 &  0.95 \\
COSMOS J095958.46+021530.8\tablenotemark{c} & 0.66 & 15.13 & 20.12 & 44.01 &  0.1 & 0.28 &  0.7 & -0.45 &  0.45 \\
COSMOS J095959.36+022458.4 & 0.57 & 13.67 & 20.42 & 43.02 & -0.6 & 0.18 &  0.2 & -0.26 &  0.62 \\
COSMOS J100005.36+023059.6 & 0.68 &  6.42 & 20.90 & 43.35 & -0.2 & 0.04 &  1.1 &  0.02 &  0.36 \\
COSMOS J100006.42+023343.4 & 0.75 &  8.91 & 20.96 & 43.31 & -0.4 & 0.11 &  0.6 & -0.52 &  0.85 \\
COSMOS J100013.33+022657.2 & 0.73 &  7.83 & 20.71 & 43.91 &  0.2 & 0.03 &  1.6 &  0.21 & -1.00 \\
COSMOS J100013.41+021400.6 & 0.94 &  4.24 & 20.76 & 43.68 & -0.3 & 0.13 &  0.6 & -0.32 & -1.00 \\
COSMOS J100024.09+014005.4 & 0.62 & 13.96 & 19.74 & 42.65 & -1.3 & 0.03 &  0.1 & -1.00 &  0.84 \\
COSMOS J100024.86+023302.7 & 0.50 &  8.73 & 21.05 & 43.14 & -0.1 & 0.28 &  0.5 & -0.45 &  0.61 \\
  SDSS J100028.28+024103.5 & 0.35 &  7.33 & 17.44 & 42.61 & -1.7 & 0.00 &  0.0 & -0.53 &  0.72 \\
COSMOS J100031.27+022819.9 & 0.93 &  2.69 & 21.56 & 44.04 &  0.4 & 0.07 &  1.5 & -0.18 &  0.73 \\
COSMOS J100034.04+024225.3 & 0.85 &  7.21 & 20.64 & 43.37 & -0.6 & 0.22 &  0.1 & -0.30 &  0.50 \\
COSMOS J100036.21+024928.9 & 0.47 &  5.69 & 18.77 & 43.03 & -1.0 & 0.25 & -0.4 &  0.20 &  0.67 \\
COSMOS J100037.99+014248.6 & 0.62 &  8.77 & 20.39 & 43.17 & -0.5 & 0.20 &  0.2 & -0.30 &  0.63 \\
COSMOS J100046.55+024412.0 & 0.22 & 28.43 & 20.42 & 42.27 & -0.3 & 0.00 &  1.3 &  1.00 &  0.33 \\
COSMOS J100047.93+014935.8 & 0.89 &  5.21 & 21.26 & 44.16 &  0.4 & 0.17 &  1.2 & -0.55 &  0.79 \\
COSMOS J100052.99+014123.8 & 0.68 &  3.95 & 21.84 & 42.77 & -0.4 & 0.14 &  0.4 & -1.00 &  0.37 \\
COSMOS J100059.45+013232.8 & 0.89 &  1.46 & 22.58 & 44.26 &  1.0 & 0.00 &  2.7 &  0.41 & -1.00 \\
COSMOS J100105.84+023041.0 & 0.70 & 10.16 & 20.64 & 43.64 & -0.1 & 0.13 &  0.8 &  0.11 &  0.39 \\
COSMOS J100124.06+024936.7 & 0.82 & 10.06 & 20.54 & 43.60 & -0.3 & 0.01 &  1.5 &  0.14 & -1.00 \\
  SDSS J100131.15+022924.8 & 0.35 &  4.95 & 18.26 & 42.91 & -1.0 & 0.05 &  0.3 & -0.67 &  0.71 \\
COSMOS J100139.10+023824.2 & 0.49 &  5.44 & 22.60 & 41.98 & -0.6 & 0.80 & -0.5 & -1.00 & -1.00 \\
COSMOS J100153.45+021152.8 & 0.48 &  7.97 & 19.31 & 43.12 & -0.7 & 0.16 &  0.1 &  1.00 & -1.00 \\
COSMOS J100209.70+023432.3 & 0.61 & 14.44 & 21.09 & 42.50 & -0.9 & 0.20 & -0.2 & -1.00 &  0.18 \\
COSMOS J100216.37+015008.2\tablenotemark{c} & 0.67 & 10.32 & 21.04 & 43.28 & -0.2 & 0.02 &  1.4 &  1.00 &  0.46 \\
COSMOS J100224.16+023107.7 & 0.67 &  6.09 & 21.68 & 43.14 & -0.1 & 0.26 &  0.5 & -0.55 &  0.85 \\
COSMOS J100231.26+022716.4 & 0.81 & 20.00 & 20.22 & 43.61 & -0.4 & 0.17 &  0.3 & -0.59 & -1.00 \\
COSMOS J100232.15+022925.6 & 0.80 &  5.89 & 20.88 & 42.78 & -1.0 & 0.28 & -0.4 & -1.00 &  0.81 \\
COSMOS J100237.09+014648.0 & 0.67 & 11.70 & 20.52 & 44.12 &  0.4 & 0.03 &  1.9 & -0.13 &  0.52 \\
COSMOS J100238.63+024743.1 & 0.82 &  1.90 & 21.89 & 42.68 & -0.7 & 0.00 &  1.0 & -1.00 & -1.00 \\
COSMOS J100240.30+020147.3 & 0.64 &  5.91 & 21.87 & 43.45 &  0.3 & 0.04 &  1.7 & -0.15 &  0.38 \\
COSMOS J100243.93+022340.7 & 0.66 &  3.82 & 21.13 & 42.49 & -1.0 & 0.16 & -0.2 & -1.00 &  0.30 \\
\enddata
\tablenotetext{a}{AGN undetected in the hard (2-10 keV) X-ray band
  have $HR=-1$, while those undetected in the soft (0.5-2 keV) X-ray
  band have $HR=1$.}
\tablenotetext{b}{AGN hosts which lack morphological data are assigned
  $b/a=-1$.}
\tablenotetext{c}{These optically dull AGN show significant
  variability in their blue emision.}
\end{deluxetable}
\end{footnotesize}

\end{document}